\documentclass[%
 reprint,
 amsmath,amssymb,
 aps,
prstab,
]{revtex4-2}

\usepackage{graphicx}
\usepackage{dcolumn}
\usepackage{bm}
\usepackage{bbm}

\usepackage{siunitx}
\usepackage{float}

\begin{document}

\preprint{APS/123-QED}

\title{Impact of local bunching factors in single-pass THz free electron lasers}

\author{X.-K. Li}
\email{xiangkun.li@desy.de}
\author{M. Krasilnikov}
\email{mikhail.krasilnikov@desy.de}
\affiliation{%
 Deutsches Elektronen-Synchrotron DESY, Platanenallee 6, 15738 Zeuthen, Germany
}%



\date{\today}

\begin{abstract}
In simulations for modern free-electron lasers (FEL), shot noise plays a crucial role. While it is inversely proportional to the number of electrons, shot noise is typically modeled using macroparticles, with their bunching factors corresponding to the bunching factors of the much larger number of electrons. For short-wavelength FELs, the macroparticles are assumed to be uniformly distributed on the scale of the resonant wavelength, since shot noise dominates the initial radiation - for instance, in the self-amplified spontaneous emission (SASE) regime. 
In this paper, we show that this assumption does not hold at longer wavelengths, particularly in the THz range, where the bunch current profile is not uniform even within the length of the resonant wavelength. Instead, the current profile dominates the initial bunching factors, which can be several orders of magnitude higher than shot noise. The slice-based bunching factors and bunching phases are derived for Gaussian distributions and compared with shot noise under the assumption that the current within each slice remains constant. Using the THz FEL at the photoinjector test facility at DESY in Zeuthen (PITZ) as a case study, the influence of the current profile has been benchmarked through simulations under very low bunch charge, where the full number of electrons can be modeled using the Genesis1.3 code. Additional simulations with the nominal working parameters of PITZ THz FEL have been compared with experimental data, indicating better agreement when the actual current profile is taken into account.

\end{abstract}

\maketitle


\section{\label{sec:level1}Introduction}

Shot noise, arising from the stochastic nature of the electrons in a bunch, is one of the most critical parameters in initiating electron distributions in FEL simulations. In the SASE regime, shot noise not only provides the seeding signal but also influences the properties of the subsequently amplified radiation fields~\cite{saldin2013physics}. In simulations, the shot noise is implemented to reproduce the same bunching factors from the typically limited number of macroparticles as would occur with much larger number of real electrons. For example, in Genesis1.3~\cite{genesis13}, this is achieved by adding mirror particles to the uniformly distributed macroparticles. These mirror particles are evenly shifted along the slice (typically one resonant wavelength) with a certain random jitter, as described in~\cite{penman1992simulation}. However, simulations using this shot noise model predicted significantly lower output energy at the undulator exit than observed in measurements on the THz FEL at PITZ. It was subsequently found that the assumption of uniform distribution within a resonant wavelength does not hold at longer wavelengths. This underestimates the initial local bunching factors, which are calculated by considering only electrons within one resonant wavelength and arising from the slopes of the current profile, by several orders of magnitude. As a result, the micro-bunching process is delayed, the FEL process develops later, and the output laser pulse inherits greater shot-to-shot instability. 

The importance of the current profile has been emphasized in previous studies~\cite{bonifacio1997rigorous, mcneil1999self, mcneil2002self,mcneil2003unified}, with the resulting enhancement of spontaneous radiation referred to as coherent spontaneous radiation (CSE). While those studies mainly discuss the impact of CSE in short-wavelength FELs, the present work focuses on its significance at longer wavelengths. Due to the much smaller ratio of the root-mean-square (RMS) bunch length to the radiation wavelength in the THz regime, the presence of initial bunching factors arising from the current profile can strongly enhance FEL performance, thus relaxing the design requirement of a THz FEL. As shown later, this also leads to a substantial improvement in the shot-to-shot stability of simulated THz pulses, including the spectra and arrival time jitter. 

In the remainder of this paper, a theoretical analysis of the local bunching factors- covering both their amplitude and phase - is presented in Section~\ref{sec:theory}. These are compared to the case where the beam current profile impact is neglected. Their dependence on the ratio of the RMS bunch length to the resonant wavelength is discussed. Section~\ref{sec:simulations} presents simulations based on the setup of the proof-of-principle experiments on the THz FEL at PITZ~\cite{krasilnikov2023first}, first benchmarked at very low bunch charge and then extended to nominal operating parameters, with and without considering the current profiles. The nonuniform initial longitudinal distributions of macro particles lead to completely different gain curves along the undulator and properties of output pulses at the undulator exit, with simulated pulse energies compared to experimental data. The conclusion is made in the last section.

\section{Analysis of local bunching factors}\label{sec:theory}

For a randomly distributed electron bunch or bunch slice, the shot noise arises from the discrete and stochastic nature of the individual electrons.
The shot noise plays a crucial role in FELs: it determines the initial radiation fields generated by electrons wiggling in the undulator, and underlies the fluctuation of output energy, the spiky structure of the spectrum, and the arrival time jitter. For electrons within a bunch slice of one radiation wavelength, the averaged radiation fields can be expressed as a sum of phasors of individual electrons 
\begin{equation}
\label{eq:shot-noise}
 \langle e^{-i\theta} \rangle = \frac{1}{n_e}\sum_{j=1}^{n_e} e^{-i\theta_j},
\end{equation}
where $\theta = kz$ is the phase, $k=2\pi/\lambda$ is the wavenumber, $\lambda$ is the resonant wavelength and $n_e$ is the number of electrons within the slice. If electrons are randomly distributed, then the absolute square of Eq.~\eqref{eq:shot-noise}, known as the \textit{form factor}, follows the negative exponential distribution~\cite{} with mean $1/n_e$, i.e.,

\begin{equation}
 \label{eq:negative-exponential-dist}
\bigl|\langle e^{-i\theta}\rangle\bigr|^{2} \sim X, 
\quad X \sim \mathrm{Exp}(\mu), \quad \mu = \frac{1}{n_e},
\end{equation}
where $X$ is distributed with probability density function (PDF)
\[ 
f_X(x) = \frac{1}{\mu} e^{-x/\mu}, \quad x \ge 0.
\]


Although it can be often approximated as uniform within one resonant wavelength, it will be shown later that the shape of the current profile cannot be ignored at longer wavelengths, which are comparable to the RMS bunch length. In this regime, the electrons within one wavelength radiate more coherently, and their bunching factors dominate the radiation growth.

For a known current profile of $f(z)$, the local bunching factor - considering one period of the radiation wavelength - at $z=z_s$ can be calculated by~\cite{mcneil2003unified}
\begin{equation}
\label{eq:local-bunching-factor}
  b_s(z_s) = \frac{\int_{z_s-\frac{\lambda}{2}}^{z_s+\frac{\lambda}{2}} f(z) e^{-ikz} dz}{\int_{z_s-\frac{\lambda}{2}}^{z_s+\frac{\lambda}{2}} f(z) dz}.
\end{equation}

The amplitude and phase of the bunching factor are calculated by the modulus and argument, i.e., $|b_s|$ and $\phi_s = \arg(b_s)$, respectively.

Considering the Gaussian distribution, $f(z) = \frac{1}{\sqrt{2\pi}\sigma_z}e^{-z^2/2\sigma_z^2}$, and taking the linear expansion, $f(z) = f(z_s)+(z-z_s)f^{\prime}(z_s)$, the local bunching factor becomes
\begin{equation}
\label{eq:local-bunching-factor-appr}
  b_s(z_s) \approx \frac{z_s}{k\sigma_z^2} \cdot \big(\sin(kz_s)-i\cos(kz_s)\big),
\end{equation}
where $\sigma_z$ is the RMS bunch length. The amplitude and phase are then given by 
\begin{equation}
 \label{eq:local-bunching-factor-appr-amp}
 |b_s(z_s)| = \frac{z_s}{k\sigma_z^2},
\end{equation}
and 
\begin{equation}
 \label{eq:local-bunching-factor-appr-phase}
 \phi_s(z_s) = \tan^{-1}\big(\cot(kz_s)\big).
\end{equation}

The radiation from a slice consists of two parts: 
the spontaneous radiation, which is proportional to the number of electrons, $n_e$, and the coherent radiation, which scales quadratically to $(n_e|b_s|)$. The ratio of coherent radiation to spontaneous radiation, $n_e|b_s|^2$, is considered a measure of their relative contributions to the initial signals. 

Starting with Eq.~\eqref{eq:local-bunching-factor-appr-amp}, straightforward algebra shows that $n_e|b_s|^2$ reaches its maximum at $z_s=\pm\sqrt{2}\sigma_z$. Moreover, if
\begin{equation}
 \label{eq:critical-condition}
    k^3\sigma_z^3<\frac{Q_b}{q_e}\sqrt{8\pi}e^{-1},
\end{equation}
then the maximum of $n_e|b_s|^2$ is more than 1, indicating that coherent emission becomes dominant. Here $Q_b$ is the bunch charge and $q_e$ is the electron charge. 

\begin{figure}[hb]
  \includegraphics[width=0.5\textwidth]{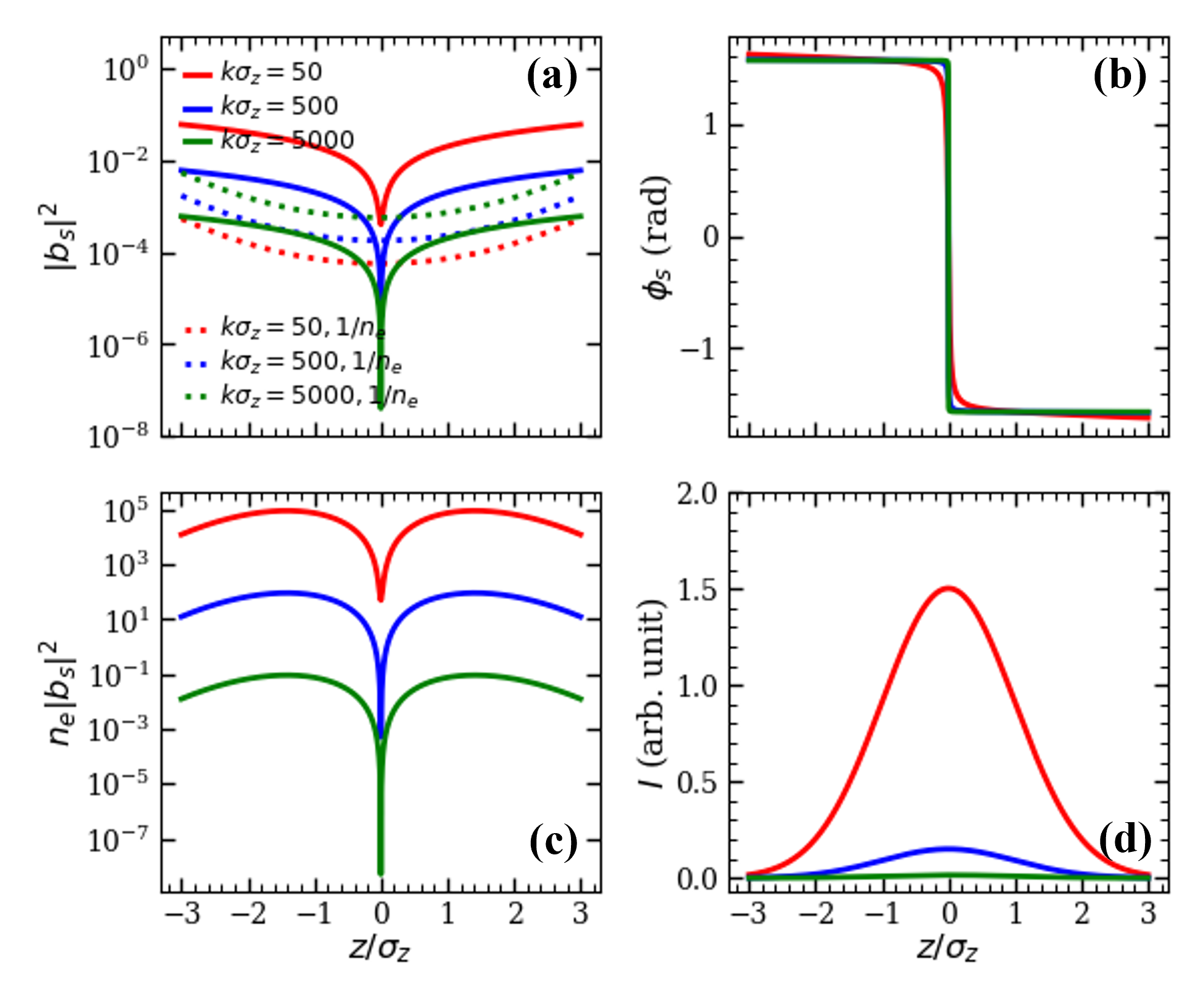}
  \caption{\label{fig:local-bunching-factor} (a) Form factors, (b) bunching phases, (c) ratios of coherent radiation over spontaneous radiation, and (d) current profiles along the bunch slices for $k\sigma_z = 50, 500, 5000$.}
\end{figure}

\subsection{Bunching amplitude and phase without noise}

In Fig.~\ref{fig:local-bunching-factor}(a) and (b), the form factor ($|b_s|^2$) and bunching phase ($\phi_s$) along slices within an electron bunch are shown for $k\sigma_z = 50, 500, 5000$, with a Gaussian current profile and a fixed bunch charge of 1 nC ($k\sigma_z<2258$ from Eq.~\eqref{eq:critical-condition}). The peak current reduces with increased $k\sigma_z$, as shown in Fig.~\ref{fig:local-bunching-factor}(d). The bunching factors were calculated using Eq.~\eqref{eq:local-bunching-factor}. It is worth noting that in a real electron bunch, the positions of electrons are intrinsically fluctuating, and there is no perfectly smooth current profile. This effect will be considered later.
The amplitude of shot noise calculated from the number of electrons in each slice, i.e., $1/n_e$, is given for comparison. For $k\sigma_z = 50$, the shot noise level (red dotted line) is more than two orders of magnitude smaller than the form factor (red solid line), therefore, coherent radiation will dominate the initial signals. This is also seen from the red curve ($n_e|b_s|^2\gg 1$) in Fig.~\ref{fig:local-bunching-factor}(c). As $k\sigma_z$ increases, the difference between the form factor and the shot noise reduces, until the shot noise dominates, as shown by the green curves ($k\sigma_z = 5000$) in Fig.~\ref{fig:local-bunching-factor}(a) and (c).
The bunching phase is almost constant along the bunch, except that it changes from $\pi/2$ to $-\pi/2$ around the peak current. This is because the real part in Eq.~\eqref{eq:local-bunching-factor-appr} is almost zero over the whole range but changes its sign at $z_s = 0$.

Consider an electron bunch with a charge of 0.1 nC and a peak current of 2.5 kA at the resonant wavelength of \SI{13.7}{nm}~\cite{ackermann2007operation, schneider2010flash}. This leads to $n_e|b_s|^2 \sim 0.1$, validating the assumption of random distribution of particles within the resonant wavelength. In contrast, consider 1 nC with a reduced peak current of 100 A lasing at 3 THz~\cite{krasilnikov2023first}, then $n_e|b_s|^2 \sim 10^4$, which implies the local current profile must be taken into account to properly represent the initial conditions in simulations.

\subsection{\label{sec:bunching-with-noise}Bunching amplitude and phase with shot-to-shot noise}

Now consider a real electron bunch with a limited number of electrons within each slice. To calculate the noise from the fluctuation of their positions, one first generates $n_m$ random samples, 
\[
z_1^*, z_2^*, \ldots, z_{n_m}^*, \quad z_i^* \sim U(0,1),
\]
where $n_m$ is the number of macro particles representing $n_e$ electrons. To reproduce the shot noise,  the samples must satisfy the condition~\cite{penman1992simulation}
\begin{equation}
|\langle e^{-j2\pi z_i^*}\rangle|^2 \sim \text{Exp}(1/n_e),
\end{equation}
for example, using the method described in Ref.~\cite{penman1992simulation}.

For each longitudinal bunch slice, the cumulative density function (CDF) 
\[
F(z)\in [0,1], \quad -\lambda/2<z-z_s<\lambda/2,
\]
is computed. The longitudinal coordinates are then obtained by
\begin{equation}
\label{eq:random-to-profile}
z_i=F^{-1}(z_i^*).
\end{equation}
In this way, one obtains not only the actual current profile but also the imprinted noise in $z^*$, which introduces fluctuations to the bunching factors.

\begin{figure}[!h]
  \includegraphics[width=0.5\textwidth]{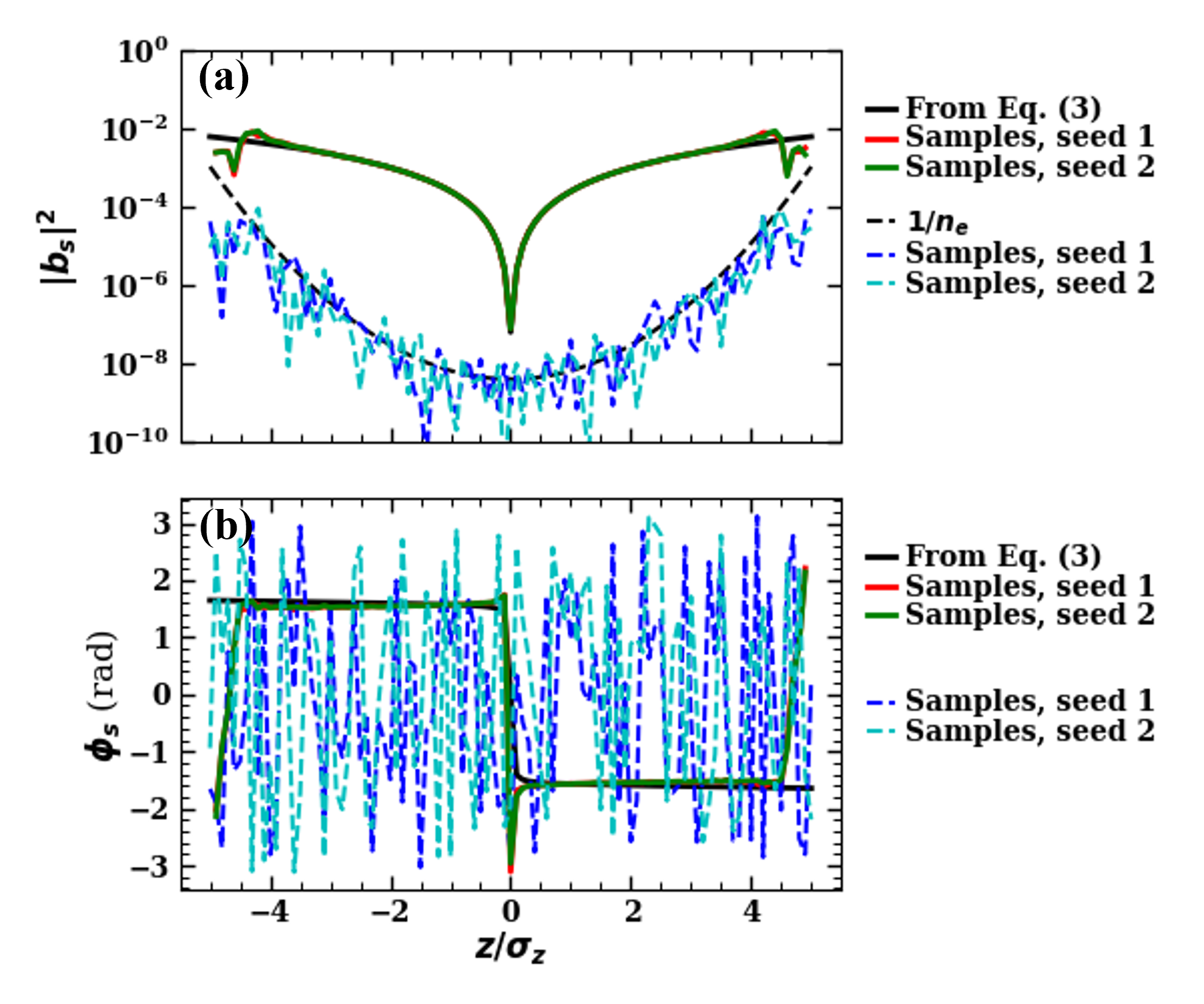}
  \caption{\label{fig:local-bunching-from-samples} (a) Square of the absolute bunching factors and (b) bunching phases along the bunch slices, with solid lines calculated from current profiles generated by random samples and dashed lines directly from random samples.}
\end{figure}

In Fig.~\ref{fig:local-bunching-from-samples}, we comapre the calculated amplitude and phase of the local bunching factors from the longitudinal coordinates ($z$) generated using the above method, with those arising from the shot noise in random samples ($z^*$) via Eq.~\eqref{eq:shot-noise}. A Gaussian distribution was assumed, with 10k macro particles per slice. As before, the bunch has a charge of 1 nC and a peak current of 100 A, and the resonant wavelength is \SI{100}{\um}.
As expected, the shot noise from the random samples ($z^*$) fluctuates around $1/n_e$, while the form factor from the longitudinal coordinates ($z$) shows excellent agreement with Eq.~\eqref{eq:local-bunching-factor-appr-amp}. Small fluctuations along the bunch slices are visible near the tails, where the distribution is nearly uniform.

\begin{figure*}[!ht]
\includegraphics[width=\textwidth]{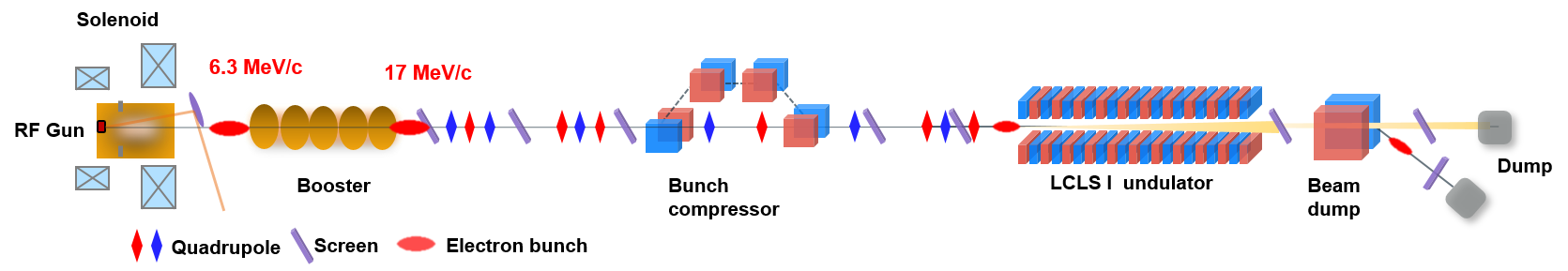}
\caption{\label{fig:THz-setup}The layout of the THz beamline at PITZ. The L-band photocathode RF gun and the booster accelerate the beam momentum to 17 MeV/c, followed by four quadupole triplets to transport and match the space charge dominated electron beam to the undulator.}
\end{figure*}

\begin{figure*}[!ht]
\includegraphics[width=0.8\textwidth]{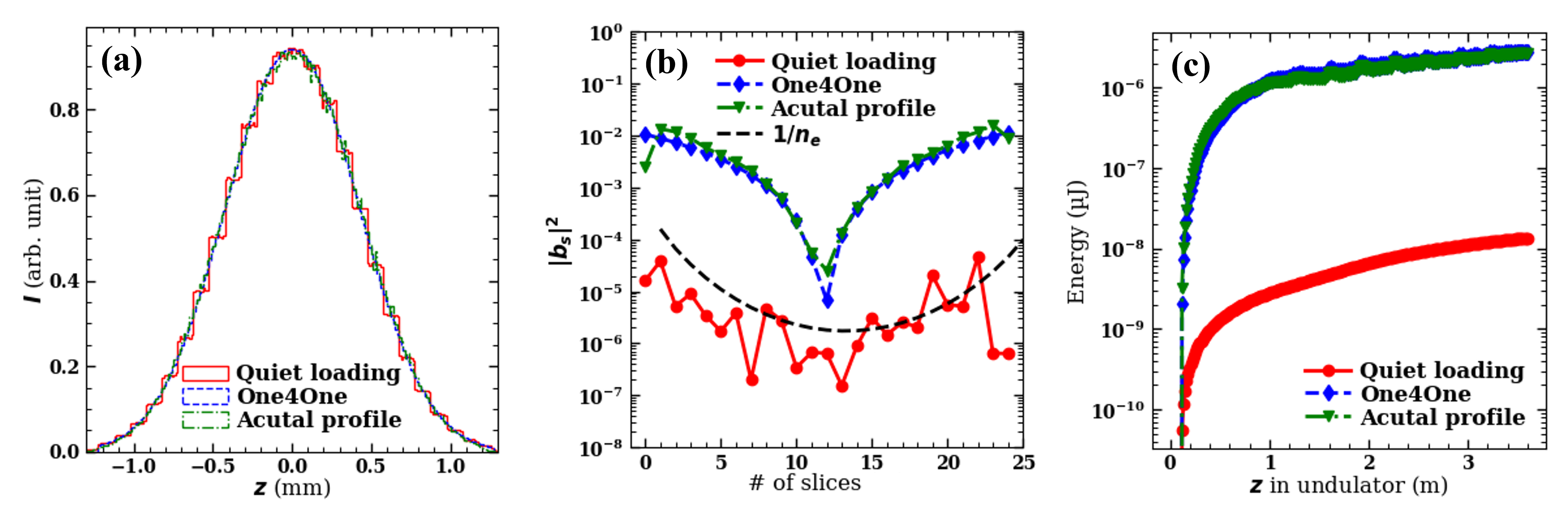}
\caption{\label{fig:Benchmark-sim} (a) Current profiles, (b) Slice bunching factors and (c) gain curves from benchmark simulations.}
\end{figure*}

The random samples exhibit a random bunching phase ranging from $-\pi$ to $\pi$, while the actual distribution generates a nearly constant bunching phase before and after the peak current, with negligible deviations along the bunch slices resulting from the noise. These results indicate that not only is the form factor significantly enhanced by the current profile, but also the bunching phases are well aligned between neighboring slices. This provides the THz FEL a start-up "seeding" signal.

\section{\label{sec:simulations}Simulations and discussion}

In this section, we simulate the THz FEL currently operating at PITZ~\cite{krasilnikov2023first} using Genesis1.3~\cite{genesis13} to illustrate how the current profile and the initial bunching factors affect the lasing process and the properties of the emitted radiation pulses. The layout of the FEL is shown in Fig.~\ref{fig:THz-setup}. The photojector comprises an L-band normal conducting photocathode RF gun and a cut-disk structure (CDS) booster accelerator, which accelerates the beam to momenta of up to 6.7 MeV/c and 22.5 MeV/c, respectively. An LCLS-I type undulator~\cite{trakhtenberg2005undulator}, on loan from SLAC, has been installed at the end of the beamline to generate THz radiation. The nominal beam momentum is 17 MeV/c, corresponding to a resonant wavelength of \SI{100}{\um} (3 THz). Four quadrupole triplets are used to transport, focus and match the space charge dominated electron beam, typically 2 nC, into the undulator. The nominal beam parameters and the undulator properties are given in Table~\ref{tab:THz-FEL}. The generated pulse energy ranged from several tens to more than \SI{100}{\uJ}~\cite{krasilnikov2025first}. 

\begin{table}[h]
\caption{\label{tab:THz-FEL} The main parameters for the PITZ THz FEL.}
\begin{ruledtabular}
\begin{tabular}{lcc}
\textrm{Parameters}&
\textrm{Values}&
\multicolumn{1}{c}{\textrm{Units}}\\
\colrule
Beam momentum & 17 & \rm{MeV/c} \\
Bunch charge  & 2  & \rm{nC} \\
Beam emittance & 4 & \rm{mm mrad} \\
Peak current & 112 & \rm{A} \\
Undulator period & 3 & \rm{cm} \\
Undulator parameter & 3.47 & \\
Resonant frequency & 3 & \rm{THz} \\
\end{tabular}
\end{ruledtabular}
\end{table}

\subsection{Benchmark simulation with low bunch charge}

To validate the numerical method in Section~\ref{sec:bunching-with-noise} used to generate input particles for Genesis1.3, we consider a beam with a charge of 1 pC and a peak current of 0.3 A. Such a beam allows us to perform the FEL simulation with the full number of electrons (one macroparticle representing one electron, \textbf{\small One4One} in Genesis1.3). For comparison, two additional simulations were carried out: (1) with \textbf{quiet loading}, where macroparticles were initialized randomly in the longitudinal coordinates with shot noise inversely proportional to the number of electrons in each slice, and (2) with the \textbf{actual profile}, where macroparticles were initialized longitudinally based on the current profile in each slice following Eq.~\eqref{eq:random-to-profile}. In the \textbf{\small One4One} case, the number of macroparticles is equal to the number of electrons (6,241,509 in total). In the other two cases, the bunch was divided into 26 slices with 32,768 macroparticles per slice (851,968 macroparticles in total). 

The corresponding current profiles from the macroparticles are shown in Fig.~\ref{fig:Benchmark-sim}(a) with a bin size of $\lambda_s/8$. In the quiet loading case, the particle distribution exhibits a step-like structure, with each step corresponding to a bunch slice, whereas in the other two cases the particle distribution is smooth. The slice bunching factors and the gain curves are shown in ~\ref{fig:Benchmark-sim}(b) and (c), respectively. In the quiet loading case, the bunching factors are nearly two to three orders of magnitude smaller than in the other two cases, which result in the much lower THz pulse energy at the undulator exit. The excellent agreement is found between the \textbf{\small One4One} case and the actual profile case highlights the critical role of the current profile in the simulations; this validates the approach of initializing the particles using Eq.~\eqref{eq:random-to-profile}.

\subsection{Simulation of the THz FEL at PITZ}

\begin{figure}[h]
  \includegraphics[width=0.25\textwidth]{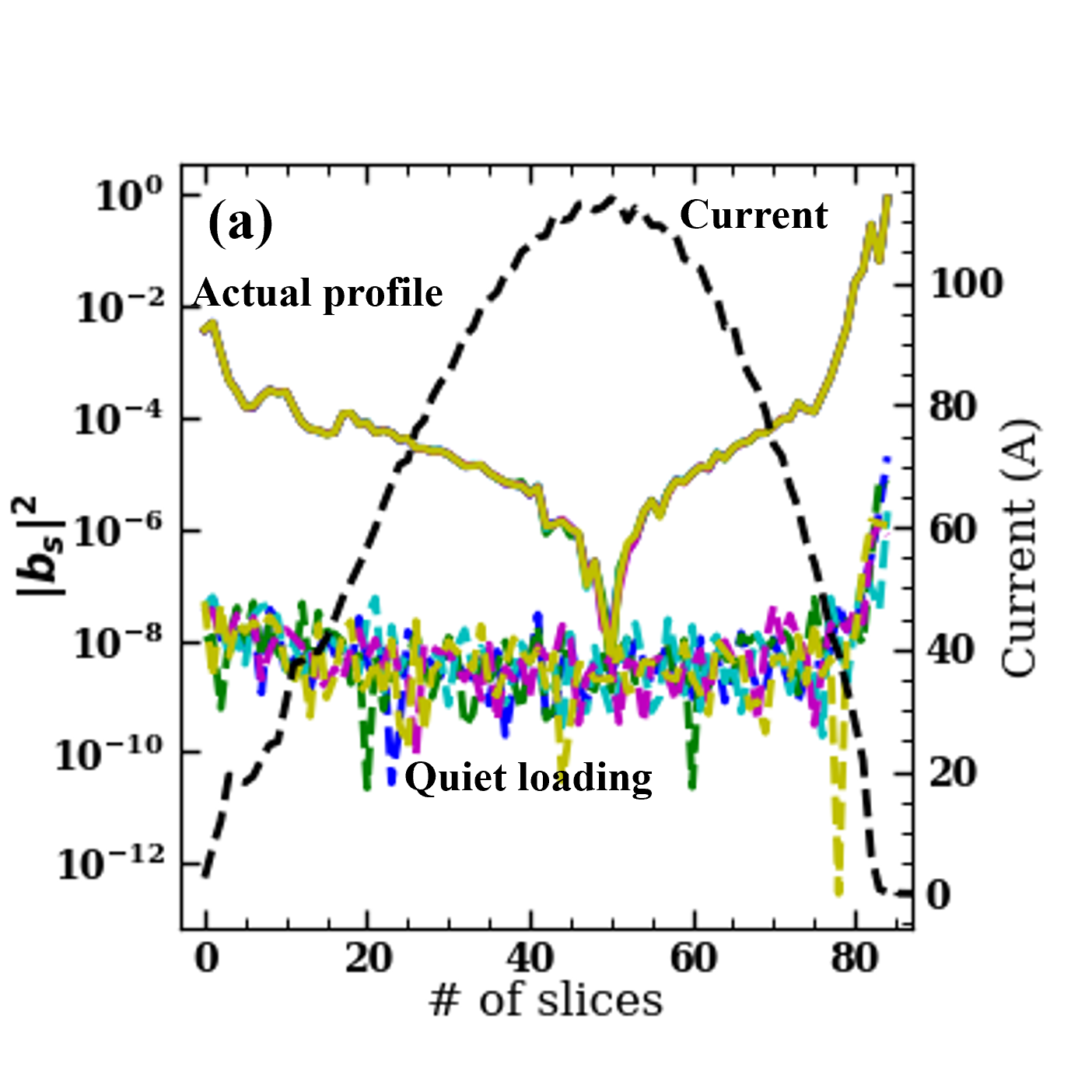}%
  \includegraphics[width=0.25\textwidth]{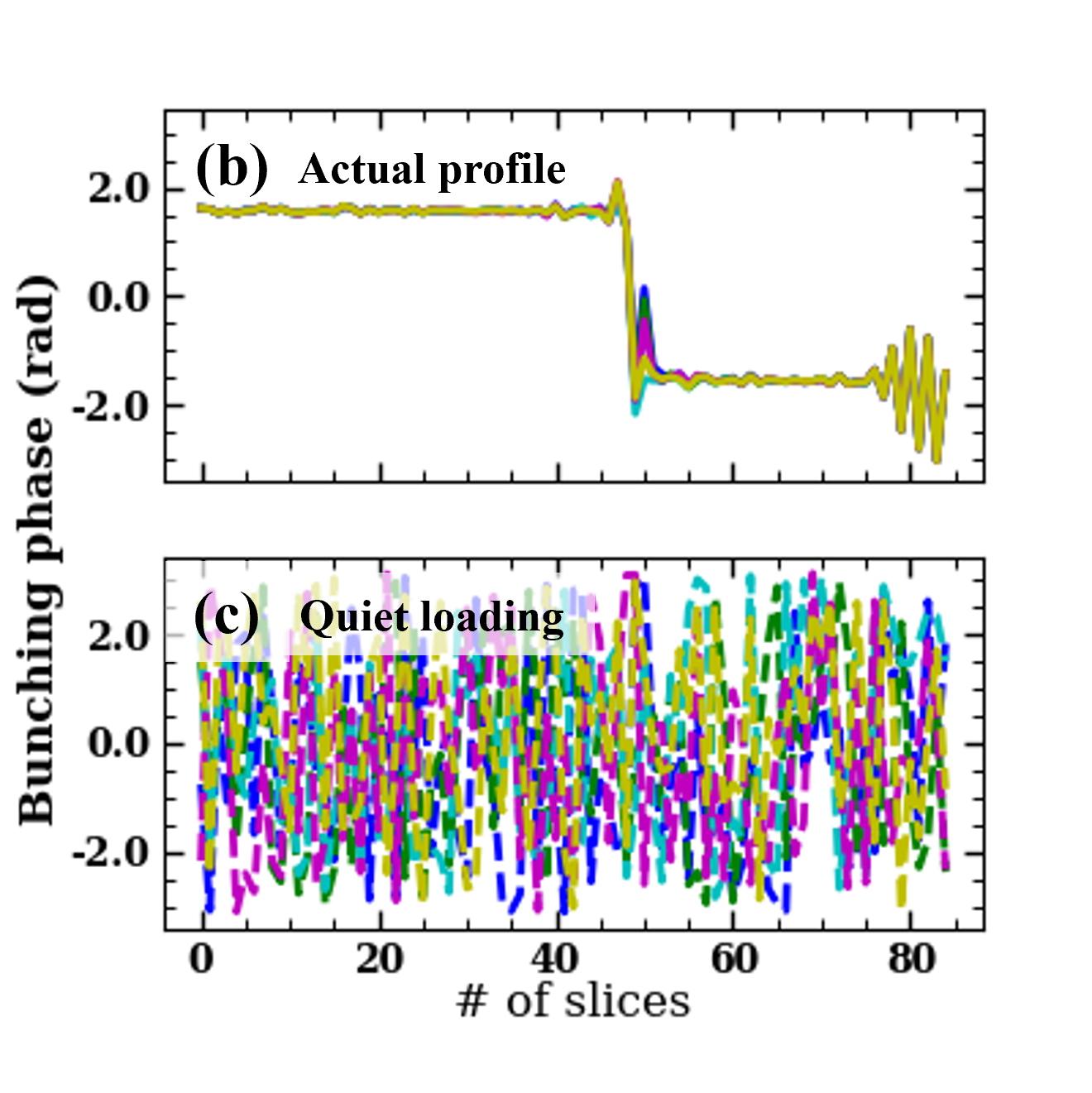}%
  \caption{\label{fig:initial-bunching}(a) Initial bunching amplitudes and (b,c) initial bunching phases in Genesis1.3 simulations with multiple numerical seeds, obtained using the actual current profile (solid curves) or quiet loading (dashed curves) in each slice.}
\end{figure}

To provide a reasonable input particle distribution for Genesis1.3 simulations and to compare with experimental data as given in Talbe~\ref{tab:THz-FEL}, start-to-end beam dynamics simulations were performed from the photoinjector to the undulator entrance, using Astra~\cite{astra} for the photoinjector and Ocelot~\cite{ocelot} for the remaining beam transport. Since the number of electrons is large, FEL simulations were carried out in two cases: (1) with quiet loading and (2) with the actual profile, as explained above. In both cases, each slice contains 32,768 macroparticles with numerical seedings taken into account. The initial bunching factors along the slices are compared, generating much stronger seeding signals (Fig.~\ref{fig:initial-bunching}(a)) and more stable bunching phases (Fig.~\ref{fig:initial-bunching}(b) and (c)) in the actual profile case than those in the quiet loading case.


\begin{figure}[!h]
  \includegraphics[width=0.32\textwidth]{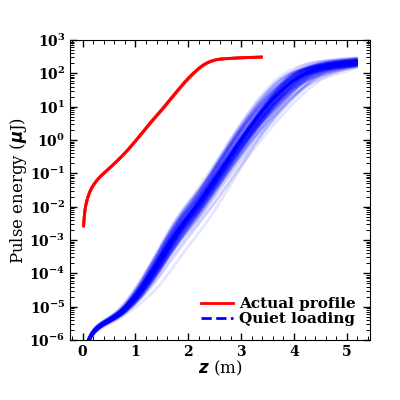}
  \caption{\label{fig:energy-vs-z} Averaged gain curves from Genesis1.3 simulations with multiple numerical seeds, applying the actual current profile (red solid) or quiet loading (blue dashed) in each slice. The light curves show individual numerical seeds, which overlap for the actual profile and are barely visible. The undulator exit is at 3.4 m.}
\end{figure}

\begin{figure}[!h]
  \includegraphics[width=0.5\textwidth]{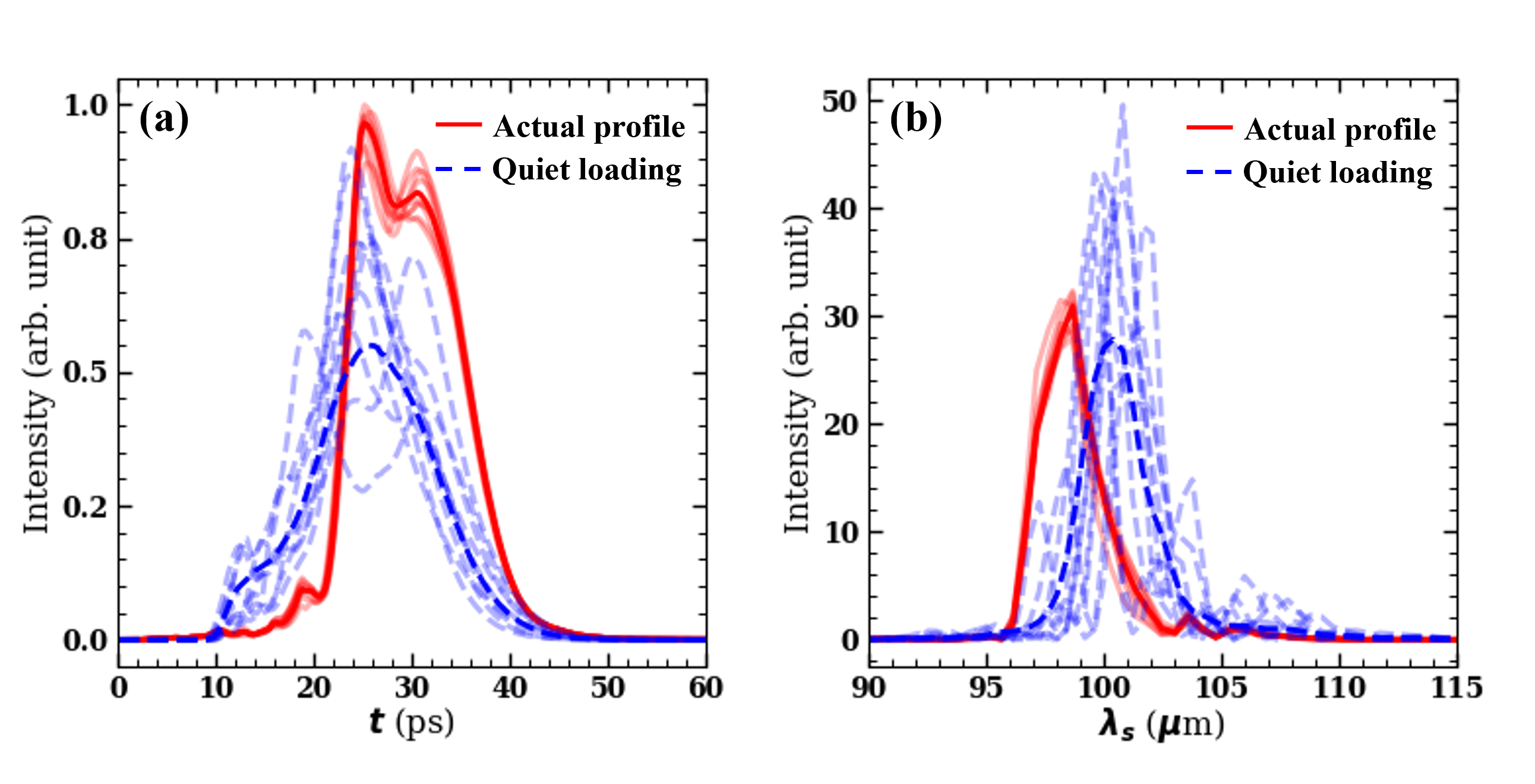}
  \caption{\label{fig:spectra-and-profiles} (a) Temporal profiles and (b) spectra of the THz pulses from Genesis1.3 simulations with multiple numerical seeds, obtained using the actual current profile (solid curves) or quiet loading (dashed curves) in each slice.}
\end{figure}

The corresponding gain curves are shown in Fig.~\ref{fig:energy-vs-z}.
In the case of the actual current profile, the initial signal had an energy of nearly three orders of magnitude higher, leading to a faster micro-bunching, an earlier start of the exponential gain and the onset of saturation at around 2.3 m. In the quiet loading case, the exponential gain came much later, and the onset of saturation was only reached at around 5 m, which is longer than the undulator (3.4 m). In both cases, we initiated the particles with 100 random seeds. The gain curves showed a strong dependence on the seed only in the quiet loading case, with a large shot-to-shot fluctuation of the saturated energy. The temporal profiles and spectra of the THz pulses are shown in Fig.~\ref{fig:spectra-and-profiles}. In the case of the actual profile, the THz distributions were nearly identical; in the quiet loading case, they differed from each other significantly. Similar results were also observed in the THz pulse spectra, with much better shot-to-shot stability when considering the actual current profile.

The properties of the THz pulses are summarized in Table~\ref{tab:THz-FEL}. The pulse energy fluctuation is less than 1\% for the simulations using the actual current profile (case I) but 15.3\% for the other (case II). The spectral width is reduced by 21\% to \SI{1.95}{\um} and the arrival time jitter is reduced by one order of magnitude, down to 100 fs. 

\begin{table}[h]
\caption{\label{tab:THz-FEL} Comparison of THz pusle properties.}
\begin{ruledtabular}

\begin{tabular}{lccr}

\textrm{Parameters}&
\textrm{Actual profile}&
\textrm{Quiet loading}&
\multicolumn{1}{c}{\textrm{Units}}\\
\colrule
Slice profile & actual & random & \\
Peak power & 24.2 $\pm$ 0.9 & 16.1 $\pm$ 3.8 & \SI{}{\MW} \\
Pulse energy  & 308.1 $\pm$ 2.3  & 214.9 $\pm$ 32.9 & \SI{}{\uJ} \\
Center wavelength & 98.9 $\pm$ 0.1 & 100.8 $\pm$ 0.6 & \SI{}{\um} \\
Spectral width & 1.9 $\pm$ 0.1 & 2.5 $\pm$ 0.5 & \SI{}{\um} \\
Pulse duration & 5.4 $\pm$ 0.04 & 6.2 $\pm$ 0.7 & \SI{}{\ps} \\
Arrival time jitter & 0.1 & 1.3 & \rm{ps} \\
\end{tabular}

\end{ruledtabular}
\end{table}

The measured pulse energy at 2 nC was $\sim$\SI{50}{\uJ}, with a fluctuation around 10\%. Considering the THz radiation transport loss of 50\% due to the open aperture of the reflection mirror in the beam pipe and the transmission efficiency of the vacuum-to-air extraction window~\cite{krasilnikov2025first}, nearly \SI{100}{\uJ} pulse energy was produced. In the simulations, the pulse energy at the undulator exit (3.4 m) was $7.00 \pm 4.21$ \SI{}{\uJ} and $308.14 \pm 2.29$ \SI{}{\uJ} for the quiet loading and the actual current profile cases, respectively. The corresponding shot-to-shot pulse energy jitter is 60\% and 0.7\%, respectively. Therefore, the experimentally produced pulse energy and energy jitter agreed much better with the simulations using the actual profile, despite the difference (a factor of three bigger pulse energy and a significantly better jitter from simulations). These difference can be explained by: the bunch charge jitter of 2\% as measured; the smearing-out of the bunching factors due to ripples in the current profile, originating from the emission process. The long beam transport from the booster exit to the undulator could degrade the beam quality significantly, due to collective effects such as space charge forces and wakefields from apertures in the beamline. The beam trajectory is also critical in the narrow chamber (11 mm $\times$ 5 mm wide, 3.4 m long) between the undulator magnets in the presence of the transverse gradient and the Earth's magnetic fields~\cite{krasilnikov2021modeling}. Recent simulations have shown a strong betatron motion of the whole bunch and a resulting inefficient beam-radiation interaction due to a mismatched trajectory in the vertical plane.





\section{Conclusion}
The initial conditions of particle distributions play a critical role in FEL simulations. To generate the correct bunching factors, a random distribution in the longitudinal coordinates is typically assumed. However, this can significantly underestimate the influence of the actual current profile at longer wavelengths, such as in the THz regime.

The reason is that the slope of the current profile can produce strong local bunching factors — peaking at $\pm\sqrt{2}\sigma_z$ for Gaussian profiles - which dominate the initial seeding signal once the condition in Eq.~\eqref{eq:critical-condition} is satisfied. This effect, in turn, accelerates the microbunching process, leading to an earlier onset of exponential gain and improved shot-to-shot stability. This mechanism can substantially relax design constraints, for example by reducing the requirement of bunch compression or the number of undulator modules.

In this paper, the THz FEL at PITZ was used to demonstrate the impact of local bunching factors in FEL simulations. We benchmarked a numerical method that introduces the actual current profile into Genesis1.3 simulations by considering a very low bunch charge. Excellent agreement was obtained with simulations using the full number of electrons (the \textbf{\small One4One} option in Genesis1.3). Additional simulations for the PITZ THz FEL also showed smaller discrepancies with experimental data when the actual current profile was included. Future studies will address the remaining differences between simulations and measurements, with a focus on fine-tuning the beam trajectory and transverse phase space in the undulator.

\section*{ACKNOWLEDGMENTS}
This work was supported by the European XFEL research and development program.

\nocite{*}

\bibliography{apssamp}

\end{document}